\documentclass[final
  ]{aipproc}

\layoutstyle{6x9}
\usepackage{graphicx}

\newcommand{\bea}{\begin{eqnarray}}
\newcommand{\eea}{\end{eqnarray}}
\newcommand{\be}{\begin{equation}}
\newcommand{\ee}{\end{equation}}

\begin{document}

\title{CHAPTER 7\\ {~} \\ {~} \\High energy emission from
galactic jets\footnote{Published by Cambridge Scientific Publishers (CSP)
in \emph{The Sun, the Stars, the Universe and General Relativity},  Cambridge 2011. ISBN 978-1-908106-12-4}} 



\author{H. R. Christiansen}{
  address={Physics Department, State University of Ceara, Av. Paranjana 1700, Fortaleza - CE, Brazil} }

\begin{abstract}
In this chapter we review some aspects of X-ray binaries, particularly those
presenting steady jets, i.e. microquasars. Because of their proximity and similarities
with active galactic nuclei (AGN), galactic jet sources are unique laboratories
to test astrophysical theories of a universal scope. Due to recent observational
progress made with the new generation of gamma-ray imaging atmospheric Cherenkov
telescopes and in view of the upcoming km3-size neutrino detectors, we focus especially
on the possible high-energy gamma radiation and neutrino emission. In connection with
this, we also comment about astrophysical jets present in young stellar objects, and we
briefly discuss similarities and differences with extragalactic AGN and gamma-ray
bursters.

\end{abstract}

\maketitle


\section{Introduction}

Astrophysical jets are collimated outflows that seem to occur wherever accretion of
matter with angular momentum in a gravitational potential takes place. The presence of
magnetic fields probably plays a major role in the formation and launching of such jets.
They are observed over a wide range of spatial scales, from AGN to protostars.

Binary systems with jets offer two great advantages as potential natural laboratories
for the investigation of astrophysical outflows: they can be found at relatively short
distances from the Solar System and the mechanisms involved in the production of the
observed phenomenology operate on relatively short time scales. A variety of microscopic
processes due to electromagnetic, weak and strong interactions in the jets of such
systems result in the production of radiation covering the whole spectrum, yielding a
unique source of information on the associated physics.  Ground-based and satellite
detectors can measure this radiation and probe the innermost regions of the sources as
well as the interaction of the jets with the environment.

There is clear evidence of the presence of relativistic leptons in galactic jets:
extended non-thermal emission and polarization can be measured in radio-wavelengths
and, in some cases, up to X-rays. The presence of an accretion disc of baryonic matter
and the detection of very high energy (VHE) emission strongly supports a hadronic jet's
content as well (e.g. \cite{hugo al 2005, romero china 2005}).

Relativistic electrons in the jets suffer synchrotron and inverse Compton losses. In
cases where strong photon fields from the donor star are present, complex processes like
electromagnetic cascading can take place (e.g. \cite{orellana al 2007, cerutti al
2009}).
In addition, the likely hadronic content of some jets should also result in energetic neutrinos
together with VHE gamma-ray production. Indeed, relativistic $pp$ and $p\gamma$
collisions give rise to very energetic pions rapidly decaying into leptons and
gamma-rays. Remarkably, muon-neutrinos above 1 TeV from this type of sources will
be detectable with ICECUBE in the near
future, opening a new era in astronomy (see e.g. \cite{nuestro prd 2006}
and Neutrino section, below).

In the understanding of VHE emitting sources, cooling and accelerating hadron processes
are crucial to evaluate the final maximum proton energies available in the acceleration
region. Considering that the fractional power of ultra-relativistic protons can be
determined by means of the most restrictive observational data, the local steady
distribution of parent pions and the resulting gamma-rays and neutrino fluxes can be
theoretically predicted \cite{nuestro mnras 2008}.
Current and upcoming experiments such as MAGIC II, VERITAS, HESS II, CTA and GLAST should
shed strong light on most of these issues in the next few years.

In what follows we review and highlight some topical issues in this fast-developing
field.

\section{Galactic X-ray binaries}

X-ray binaries (XRBs) contain a stellar mass compact object (CO) (white dwarfs have been
excluded) which is supposed to emit X-rays as a consequence of gravitational accretion
of matter from the companion. The companion, also known as donor, is a non-degenerated
star which can be in any stage of its evolution.

As of 2007, the hitherto last catalogue of low (donor) mass XRB (LMXB) in the galaxy signals the
identification of 185 of such systems (excluding quiescent systems) \cite{liu al 2007}.
The related catalogue for high mass XRBs (HMXB) indicates 114 of these binaries
\cite{liu al 2006}. LMXB are formed by a black hole or neutron star with a late-type
main sequence star (A, F, G) or even a white dwarf. Typical LMXB secondaries have masses
below $1 M_\odot$ and are Roche-lobe deformed by the compact object. HMXB comprise an OB
donor (Be or SG type) and a black hole or neutron star. Note however that black holes
and Be stars are not found together.

About 20\% of the three hundred XRBs catalogued, 56 LMXB and 9 HMXBs, are seen as
clear radio emitting
sources of which at least 15, and possibly up to 20, can be already enlisted in the
microquasar category.
Microquasars are rapidly variable X-ray binaries that present extended jets
strongly emitting in radio and, in some cases, up to X-rays and even higher energies.
It is supposed that the compact object
magnetically powers a relativistic jet via accretion of matter from the companion.

So far, the only well established VHE (TeV) gamma-ray emitters are 4 galactic HMXBs (PSR
B1259-63, LS I+61 303, LS 5039, Cyg X-1), the first of which harbors a non-accreting
pulsar while the last one holds an accreting black hole. The nature of the compact components
of the other two systems is not completely established since their masses are just roughly
known (1 - 5 $M_\odot$), but they are likely to belong to the microquasar category. 
Note also that there is an equivalent number of already confirmed XRBs (particularly
HMXB) in the Magellanic Clouds that might be similar to the galactic ones in every
sense. In any case, the number of objects of each class mentioned above will,
presumably, be increasing as fast as new technologies get incorporated among the
detection facilities.


\subsection{X-ray signature, accretion and spectral states}

Most of the present understanding
about jets in XRBs comes from the study of black hole (BH) candidates. The main reason
is that they are more easily detected since, in general, BH XRBs are more radio loud
than neutron star (NS) XRBs.

Accreting black holes though emit most of their luminosity in the X-ray band,
which strongly varies depending on the accretion state of the source.
Indeed, the same source can exhibit very different X-ray spectra characterizing a
diversity of XRB spectral states. There are two main such states relatively steady and
frequently observed.
At high luminosities (above $\sim$0.1 of the Eddington luminosity \footnote{$L_{\rm Edd}
\simeq 10^{38}M/M_\odot$ erg/s, where $M$ is the mass o the accreting object.}), it is
said that the accretion flow is in the High Soft State (HSS), characterized by a strong
(\textit{high}) thermal disc radiation and some reflection contributions likely associated with
 illumination between the accretion disc and the corona. It also presents a weak
(\textit{soft}) non-thermal (power-law) component extending up to high energies (X-rays
and low energy $\gamma$-rays).
This power-law, with a steep photon index $\sim$2.3-2.5, is interpreted as coming from
inverse Compton (IC) upscattering of UV and soft X-ray photons by thermal
($kT\simeq 30-50$ keV) and non-thermal
(index $\sim$3.5-4.5) electrons of a hot plasma or corona (see e.g.
\cite{DelSanto-Malzac et al 2008}). This non-thermal comptonization on top of a black
body spectrum peaking around 1 keV strongly characterizes the HSS.

At \textit{low} luminosities (below a few percent of  $L_{\rm Edd}$) the sources are mostly
found in the so-called Low Hard State (LHS) in which disc blackbody features and
reflection properties are weak. There seems to be a hot corona dominating the luminosity
output of the system and emitting a thermal distribution of electrons that boost
(\textit{harden}) the energy of X-rays \cite{malzac-belmont procs foca 2008}. LHS
spectra are very well fitted by multiple Compton up-scattering of soft photons by a
Maxwellian distribution of electrons (see \cite{sunyaev-titarchuck1980}) in a hot plasma
($kT\simeq 50-100$ keV). It is this thermally comptonized spectrum, dominated by a hard
power-law (index $\sim$1.5-1.9) with an exponential cutoff around 100 keV, that
characterizes the LHS.

Besides the LHS and HSS, there are several other intermediate states
(IMS) more difficult to define, often appearing when the source is about switching
between the two main states.
For example, the rapid X-ray transition from hard to soft states is associated with
radio flares which show optically thin spectra in a steep power-law state also known as
very high (VHS). These flares are signatures of powerful ejection events, spatially
resolved as large-scale extended jets.
Intermediate states are easily identified in hardness-intensity diagrams (HID), which
present a typical hysteresis or q-like form along the different spectral states. They
have been used with great success to study the evolution of outbursts and to distinguish
the different accretion states of a source (e.g. \cite{homan et al 2001, fender al 2004,
belloni et al 2005}), in particular the radio-loud and radio-quiet phases. In addition,
the timing properties of the X-ray light curve change dramatically with the position in
the HID (e.g. \cite{homan et al 2001, kording-fender 2006}).

The different spectral states are usually understood in terms of changes in the geometry
of the accretion flow. The standard picture (e.g. \cite{done et al 2007}) is that in the
HSS there is a geometrically thin disc extending down to the last stable orbit. This
disc is responsible for the (dominant) thermal emission and is the source of soft
photons for later comptonisation in small active coronal regions located near the disc.
There, electrons absorb energy through magnetic reconnection/dissipation processes
\cite{galeev et al 1979}. These electrons thereafter lose energy by boosting the
soft photons coming from the disc. This produces a high energy non-thermal emission which, in
turn, illuminates the disc inducing reflection features in the source \cite{Zdziarski et al
2002}.
%
When the system is steadily in the soft state, there is a quenching
in radio emission which may be due to the physical suppression of the
collimated outflow \cite{fender al 1999}.

In the standard model of spectral states, the LHS is explained by means of a truncated
geometrically thin disc \cite{Shakura et al 1973}. Instead of extending down to the last
stable orbit, this disc is terminated at large distances that range from a few hundred to a few
thousand gravitational radii from the black hole as suggested by the weakness of its
thermal features. In the inner parts, the accretion flow takes the form of a hot
geometrically thick, optically thin, corona-like structure \cite{esin et al 1997, poutanen et al 1997},
possibly advection dominated and radiatively inefficient \cite{blanford-begelman 1999}.
The electrons have a thermal distribution that can cool down by comptonisation of the soft
photons coming from the external geometrically thin disc and IR-optical photons
internally generated through self-absorbed synchrotron radiation. See also \cite{markoff
et al 2005, maccarone 2005}.
%
%
In the LHS, the observed radio emission is optically thick with a flat or slightly
inverted spectrum, and indirect evidence indicates that this is the signature of a
continuous outflow or compact jet \cite{fender 2001}.


For BH XRBs in the hard state, it has been found a non-linear correlation between the
radio and the X-ray luminosities. It runs over more than three orders of magnitude in
the X-ray band and reads $L_R\propto L_X^{0.7}$ \cite{corbel et al 2003, gallo et al
2003}. Remarkably, extending this correlation to supermassive BHs,
 there is evidence that a single 3D power-law function can fit all the BH data
(XRBs and AGN) for a given X-ray luminosity, radio luminosity and CO mass. This function
takes the approximate form $L_R \propto L_X^{0.6} M^{0.8}$ where $M$ is the mass of the
compact object \cite{merloni et al 2003,falcke et al 2004}. The existence of this
relation, connecting BH XRBs and AGN, signals that the same physical processes could be
responsible for the disc-jet coupling, irrespective of the mass of the BH involved
\cite{mirabel 2009}. In a separate section, we shall come again  to the description of the
spectral relations in the case of NS XRBs.

\section{Galactic jets}


The study of XRBs at different wavelengths has shown that a significant fraction of the
accretion power may be released in the form of radiatively inefficient collimated
outflows or jets. In general, jets are a very common feature associated with accretion
onto relativistic compact objects on all mass scales, from neutron stars  and
stellar mass black holes in XRB systems to supermassive BHs in active galactic
nuclei. Strong jets are believed to be even at the origin of gamma-ray bursts,
the most overwhelming
transient events ever detected, but seen as well among young stellar systems
(see separate section below). The investigation of the link between
the jet emission and the different accretion regimes of an astrophysical system is an
important issue which can explain much on its internal dynamics.


Galactic jets from binary systems are particularly interesting because they develop
completely in short time scales and, in many cases, we are able to survey the whole
cycle of their existence. During the LHS, radio emission from XRBs points towards the
presence of a persistent jet in some of these systems, and transient ejections also seem
to take place when spectral states switch from the LHS to the HSS.

Depending on the object, the nonthermal radiation produced in XRB jets has been resolved
in radio at very different spatial scales but also in X-rays at large scales (see e.g.
\cite{corbel al 2002}). This is clear evidence that particle acceleration takes place
in different locations of XRB jets under very different conditions (for a discussion
 see e.g. \cite{bosch 2007}).

\subsection{Launching and powering }\label{sec:launching}


Although a complete theory of jet generation is still lacking, several studies of jet
powering, acceleration, and collimation have been carried out during the last few
decades (e.g. \cite{ferreira al 2006, barkov-komissarov 2008}).
Due to the observed correlation between accretion and jet activity in this kind of
system (see e.g. \cite{fender al 2004}), a widely accepted scenario suggests that jets
are powered and fed by gravitationally driven accretion. Under certain conditions,
magneto-centrifugal forces make the matter of the accretion disc move following ordered
magnetic field lines that thread the inner regions of the disc.


As explained in \cite{spruit 2009}, outflows are produced by magnetic field lines
anchored in the material of a rapidly rotating object. Indeed, numerical simulations
show that an ordered magnetic field near the central object operates upon the ejection of
powerful outflows.

\begin{figure}
  \includegraphics[height=.2\textheight]{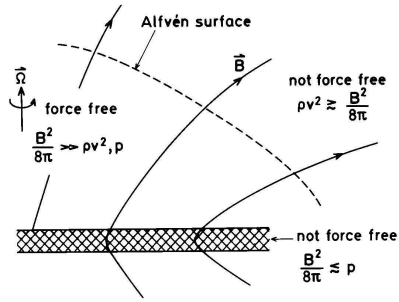}\label{spruit1}
  \caption{Regions in a magnetically accelerated flow from an accretion disc. The central
object is assumed at the left of the sketch.  Figure from \cite{spruit 2009}.}
\end{figure}

In the standard magneto-centrifugal acceleration model for jets produced by an accretion
disc \cite{BKR 1976, blandford-payne 1982} three different regions can be distinguished
\cite{spruit 2009}:
\textit{(i)} The accretion disc itself, where the kinetic energy density dominates over
the magnetic one  and makes the field lines corotate with the disc. \textit{(ii)} An
atmosphere of plasma extending outside the disc which, assuming it is cool, has a low
density and pressure. The gas is therefore dominated by the magnetic pressure which then
drives the flow into corotation. The velocity component along the field is though
unrestricted and the centrifugal force accelerates the gas along the field lines.
Consequently, bulk-matter loading and acceleration depend on the inclination of the
field lines and a net upward force acts only if they are sufficiently inclined outward
 \cite{blandford-payne 1982}. {It implies that the conditions
for acceleration and collimation (i.e. the degree to which the jet flow lines are
parallel) are rather opposed so that highly collimated jets would require further
explanation.} In addition, note that for a given field shape the mass loading decreases
with increasing field strength \cite{ogilvie-livio 2001}.
\textit{(iii)} Near the Alfvén radius (i.e. the distance from
the CO where the \textit{ram} pressure equals the magnetic pressure) the rigid
corotation approximation breaks down.  While the flow accelerates with the increasing
inclination of the field lines, the field strength decreases with
distance (see Model section).
The velocity of the bulk is very high and the matter flow starts
separating from the field lines. Its rotation rate gradually vanishes
in order to conserve angular momentum as the flow continues to expand away from the
axis. On the other hand, in this region the field lines
become nearly azimuthal, at least in a pure axisymmetric model.


After launching, the flow is first accelerated  by the centrifugal effect up to a
distance of the order of the Alfvén radius.  When the temperature in this region is
high, for example in the presence of a hot corona, the atmosphere extends higher on and
it is easier for the  mass flow to get started. If the disc atmosphere is cool
(well below the virial temperature), the gas density declines rapidly
with height and the mass flow rate becomes more sensitive to the physical conditions
near the disc surface. In the radial direction, the flow velocity increases almost
linearly with distance (from the rotation axis) up to the magnetosonic speed
($\simeq$ sound speed $c_s$), 
and thus the  mass flow rate attains the value $\dot m
\simeq c_s\rho_s$, where $\rho_s$ is the gas density at the maximal velocity.

\begin{figure}
  \includegraphics[height=.2\textheight]{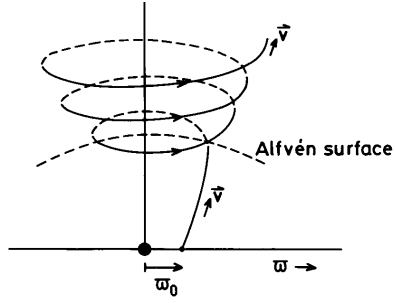}\label{spruit2}
  \caption{Magnetic field lines and particle path in the outflow. $\bar\omega$
is the orthogonal distance from the compact object. Figure from \cite{spruit 2009}.}
\end{figure}

\vskip 0.5cm

The matter contents of the jets are not exactly known, but the presence of relativistic
hadrons in the jets of SS 433 has been directly inferred from iron X-ray line
observations (e.g. \cite{kotani al 1994, kotani al 1996, migliari al 2002}). In
addition, the large deformations that some jets induce in the interstellar medium
suggest a significant baryon load \cite{gallo al 2005, heinz 2006}. The fact that the
jets are usually well collimated also favors this interpretation since cold protons help
to provide confinement to the relativistic gas.

Jets powered by the rotation of a black hole (Blandford-Znajek mechanism) would
presumably consist of a pure electron-positron pair plasma, while outflows from rotating
discs are regarded as consisting of normal hadronic plus leptonic matter. However, since
isolated black holes cannot sustain a magnetic field, a field threading the hole requires
an accretion disc to keep it in place.
If the accretion disc is cool, the conditions for outflow are sensitive to the field strength
and inclination near the disc surface. But this dependence is attenuated in an
ion-supported flow \cite{rees al 1982}, where the temperature of the hadronic matter is
near the virial temperature and the flow only weakly bonds in the gravitational potential
of the accreting object. This may, in part, be the reason why powerful jets tend to be
associated with the hard states in X-ray binaries for which the ion-supported flow (also
called ADAF) is a likely model \cite{chen al 1995}.
Therefore,  it is quite probable that a jet accelerated
by the hole is also supplied with ions besides electrons from the disc, rather than with
a pure lepton pair plasma generated after quantum fluctuations and gamma-ray annihilation.

This possibility also suggests that the magnetic flux passing through
a disc is a global quantity of the system rather than a function of local
conditions near the center. Namely, the flux depends on the way it is
transported through the disc as a whole.
In fact, some similarities in the features and time variability of spectral states of very
different brightness (c.f. VHS and LHS) \cite{rutledge al 1999, belloni al 2005,
homan-belloni 2005}, support the idea that the phenomenology of XRBs is not simply a
function of the instantaneous accretion rate alone. A useful second parameter would
therefore be a global quantity, a property of the disc as a body, which can vary among
discs or with time in a given disc. As argued in \cite{spruit-uzdensky 2005}, a
promising candidate is the net flux of magnetic field lines crossing the disc
since its value is determined only by inheritance from the
initial conditions and the boundary conditions at its outer edge. In addition,
it cannot be
changed by local processes within the disc.
An interesting observational connection is that the presence or quenching of jets could
be related to variations in the global magnetic flux of the disc \cite{spruit-uzdensky
2005}.
If the hard X-ray state is indeed one with a high magnetic flux in the inner disc, its
connection with jets would be natural since the current theory strongly suggests that
outflows are magnetically driven phenomena.

\vskip 0.3 cm

Although axisymmetry has been the standard working hypothesis,  the energy carried in
the form of the wound-up magnetic field can decay by internal dissipation in ways that
do not occur in axisymmetric systems. Indeed, a way of dissipating magnetic energy is to
generate the flow from a non-axisymmetric rotating magnetic field. A classic example of
such a flow is the pulsar wind generated by a rotating neutron star with a magnetic
field inclined with respect to the rotation axis. Thus a non-axisymmetric rotating
magnetic field turns out to be a very efficient way of accelerating the flow to high
Lorentz factors \cite{drenkhahn 2002, giannios-spruit 2006}.

At short length scales, dissipation of magnetic energy by reconnection of field lines
can be very efficient and have striking effects on the flow. For example, the conversion of
Poynting flux into kinetic energy by wasting magnetic energy. This mechanism does not
require an increasing opening angle of the flow lines. If the dissipation is a consequence of
magnetic instabilities, it works much better at high degrees of collimation. If it is due to
reconnection in an intrinsically non-axisymmetric flow,  it works independently of the
degree of collimation \cite{drenkhahn 2002}. The mechanism is also effective at
distances significantly beyond the Alfvén radius.

\subsection{Jet's scale sectors and particle interactions}


The energy release powering a relativistic outflow takes place near the black hole, at a
distance of the order of 50$R_g$ (about 10$^{7}$ cm in the case of a microquasar). On the
other hand, narrow jets seen at radio wavelengths may appear on scales up to ten orders
of magnitude above that. The collimation of the flow, however, may take place on
intermediate scales,
large compared with the Alfvén radius, at least for very narrow jets. Such intermediate
length scales can also be crucial for acceleration to high Lorentz factors, although the
region around the Alfvén radius plays the main role in standard axisymmetric centrifugal
acceleration processes.


Once the jet is launched, it may interact with other material coming from the accretion
disc or the stellar companion. For instance, an accretion disc wind can enhance
 collimation and stability but a strong stellar wind may lead
to jet bending and even disruption into separate jet clumps.

In any case, interaction with the environment may induce shock formation and its radiative
counterpart may be observable either as a transient phenomenon, when the outflow
penetrates for the first time through the surrounding medium, or as a steady one when the jet
feeding is continuous, thus allowing re-collimation shocks.

Although the jet's environments at large scales may be quite different among microquasars,
depending on the local ISM of the galaxy and the strength of the companion's wind,
jets must terminate abruptly or be softly diffused away.
In the first case it is expected that they get stopped against the ISM by disruption
or via shocks, inducing different radiative outcomes. A classical
example of the interaction between a XRB jet and its environment is the case of
SNR W50, where the jets of the microquasar SS 433 have strongly deformed the
remnants up to the degree scale (e.g. \cite{brinkmann al 2007}).

In order to describe the jet, we can divide it into four scale regions as explained in
\cite{bosch-khangulyan 2009}. The jet base region, close to the compact object (50 -
1000$R_{g}$), the binary system scale region (typically $10^{10} - 10^{13}$ cm in XRBs),
a third one at middle scales (around $10^{15} - 10^{16}$ cm), and the termination region
of the jet, where it ends interacting somehow with the interstellar medium ($\geq
10^{17}$ cm for XRBs).

Actually, at the base the jet could be further subdivided in two parts: a close,
magnetically dominated sector, where particle acceleration may happen by magnetic energy
dissipation via magneto-hidrodynamical (MHD) instabilities (as it happens in
extragalactic jets  \cite{zenitani-hoshino 2001}), and a shock sector where the
magnetic field is in subequipartition and acceleration likely occurs via Fermi processes.
A magneto-centrifugal mechanism also operates very close to the rotating object (see
e.g. \cite{neronov-aha 2007,rieger-aha 2008}). If jet velocities are high enough at the
base, the dense available photon and matter fields from accretion disc and corona could
allow the converter mechanism to take place (see e.g. \cite{derishev al 2003, stern-pou
2006}). Magnetic field reconnection in the surrounding corona can also inject a
nonthermal population of particles into the jet enriching the internal dynamics (see e.g. \cite{gierli-done 2003} and references therein).

As we have already seen, we can assume that the presence of both leptonic and hadronic
matter in the jet is a natural consequence of its accretion origin. The relevant
radiative mechanisms at the base are leptonic synchrotron emission (see e.g.
\cite{markoff et al 2001}), relativistic Bremsstrahlung from electrons interacting with
jet ions, synchrotron self-Compton (SSC) (see e.g. \cite{bosch-paredes 2004}) and inverse Compton scattering  with
corona and disc photons (see e.g. \cite{romero al 2002, georga al 2002}), all of which
depend on the dominant local energy balance. Regarding hadronic processes, there are
several mechanisms that could produce high energy radiation in the form of gamma rays
and neutrinos, and as a by-product, low-energy emission from secondary particles. Two of
these mechanisms are collisions of relativistic jet protons against cold ions (p-p scattering),
and with photons (photo-meson production). Ions and multi-wavelength photons can have any
origin, from the accretion disc or the external medium.
Relativistic proton collisions with ions and photons
produce neutral pions  that decay into gamma rays as well as charged pions that decay
into muons and neutrinos (going in turn into electrons and other neutrinos). Another possible
hadronic mechanism is photo-disintegration, which requires the presence of ultra high
energy heavy nuclei and a dense field of target photons of large-enough energy. This
process produces lower-mass hadrons and gamma rays.

In the second jet sector, at binary system scales, plausible mechanisms for generating
relativistic particles in the jet are the Fermi processes: shock diffusive (Fermi I),
random scattering (Fermi II), and shear acceleration (see e.g. \cite{protheroe 1999}).
The Fermi I mechanism could take place due to internal shocks in the jet; Fermi II
acceleration could take place if magnetic turbulence is strong enough, with high Alfvén
velocity; shear layers would be a natural outcome in laterally expanding jets or any time
different jet vs. medium velocities take place. Interactions with the stellar wind may
also trigger particle
acceleration through a recollimation shock formed in the jet that expands against the dense
material expelled by the companion star (see e.g. \cite{perucho-bosch 2008}). The
velocities of the different shocks could be either mildly or strongly relativistic. In
the latter case, the converter mechanism may be effective in very bright star systems.
At such binary system scales, possible radiative leptonic processes are synchrotron
emission (see e.g. \cite{paredes al 2006}), relativistic Bremsstrahlung (see e.g.
\cite{bosch al 2006}), SSC (see e.g. \cite{atoyan-aha 1999,dermer-bottcher 2006}), and
external IC (see e.g. \cite{dermer-bottcher 2006, paredes al 2002, kaufman al 2002,
khangulyan al 2008}). However, jet proton collisions with target nuclei
of the stellar wind (see e.g. \cite{romero al 2003}) seem to be the most efficient
hadronic process at these scales. As we shall further discuss, this mechanism would lead as well to
neutrino production (see e.g. \cite{levinson-waxman 2001, nuestro prd 2006}). Other
 hadronic processes expected in this sector are photomeson production (see e.g. \cite{aha 2006}) and
photodisintegration (see e.g. \cite{bednarek 2005}).

At middle scales, intermittent ejections at larger time scales (hours to days) could
create shocks \cite{bosch-khangulyan 2009}.
Additionally, Fermi-II type and shear acceleration could drive a continuous
outflow at these scales (c.f intra-knot regions of extragalactic jets; e.g. \cite{rieger
al 2007}). Given the high jet/medium relative ram pressure at these scales, the
environment influence is not expected to be significant. Here, the emission is usually
characterized by synchrotron radiation; stellar IC upscattering is quite inefficient
because of the large distances to the companion star rarify star photons. Nevertheless,
for powerful ejections, SSC could still be significant (see e.g. \cite{atoyan-aha
1999}). Regarding the particle energy distribution, its evolution is likely dominated by
convective and adiabatic energy losses \cite{vanderlaan 1966}.

At the jet termination region, the inertia of the interstellar external medium plays an
important dynamical role (c.f. AGN hot spots and radio lobes, e.g. \cite{kaiser-alex
1997}). When the outflow hits the interface with the external plasma, two shocks may be
formed; one moving backward into the jet and another moving forward (the bow shock).
Fermi-I type acceleration mechanism then seems to be plausible, although high diffusive
and convective rates in the downstream regions of both the forward and reverse shocks
could prevent efficient acceleration. It is also possible that MHD instabilities mix the
jet matter with the interstellar medium without forming strong shocks (see e.g.
\cite{heinz-sunyaev 2002}).
If, on the other hand, particle acceleration and confinement
were efficient,
 then synchrotron, relativistic Bremsstrahlung, and IC radiation
could be produced there and eventually detected from nearby sources (see e.g.
\cite{bordas al 2008}). Hadronic acceleration could take place as well, which could lead
to lower energy production and secondary leptonic emission.

\subsection{Jets from neutron stars}

Some particular features make it worth addressing separately the issue of neutron stars
in the context of XRBs.
Recently, the Spitzer Space Telescope opened a new observational window for the study of
jets in low-luminosity neutron star X-ray binaries. Observations of the NS ultracompact
XRB 4U 0614+091 revealed the first results from the follow-up multi-wavelength
observations in the radio band (VLA), mid-IR/IR (Spitzer/MIPS and IRAC), near-IR
(SMARTS), optical-UV (Swift/UVOT), soft and hard X-rays (Swift/XRT and RXTE), the best
coverage of a NS XRB to date \cite{migliari 2008}.

With the present data, sometimes it is possible to perform an estimate of
both the total jet power and the fraction
of the accreted power channeled into the jet. Moreover, observations have generally
shown evidence for an association between the formation of a jet and an X-ray state
transition (c.f. the NS XRB GX 17+2).

XRBs holding a neutron star show some common features with BH XRBs. Remarkably, the
presence of steady jets in the low state (below a few percent of $L_{\rm Edd}$) show a
correlation between $L_X$ and $L_R$. On the other hand, the occurrence of transient
outflows at high luminosities (near $L_{\rm Edd}$), and even rapid X-ray state changes have
also been observed. All this indicate a coupling between the jet and the inner accretion
disc which does not depend on the nature of the compact object. Nevertheless, besides the
similarities, some differences deserve attention. Particularly,  the steeper relation between
radio and X-ray luminosities, $L_R\propto L_X^{1.4}$, signaling that neutron stars
remain less radio-loud than BH for a given X-ray power. 

In order to compare NS and BH samples, it is important to know the common value of $L_X$
at which it should be done. As argued in \cite{migliari-fender 2006}, the correlation
between radio luminosity and jet power $L_J$ should be compared at the least radiatively
inefficient point, while still in the hard state, therefore producing a steady jet in
both samples. This point corresponds to the brightest LHS / LAS (low hard atoll states).
Comparison of the fits to the NS and BH samples indicates that at $L_X \sim
0.02L_{\rm Edd}$, the ratio of radio luminosities is $\sim$ 30. As noted in \cite{migliari
al 2004}, assuming a scaling $L_R \propto L_J^{1.4}$ this indicates that neutron star
jets are about one order of magnitude less powerful than black hole jets at this X-ray
luminosity. This implies that $L_J \propto L_X^{0.5}$ for BHs and $L_J \propto L_X^{}$
for NS \cite{migliari-fender 2006}. In the first case, it is indicative of a radiatively
inefficient accretion, in the sense that most of the liberated gravitational potential
is carried kinetically in the bulk flow and not radiated.

Assuming that the relation between $L_J$ and the mass accretion rate of the CO,
$\dot m$, is the same for BH and NS XRBs,
Migliari and Fender show that $L_X\sim \dot m$ for NS and $L_X \sim \dot m^2$ for BHs.
The different coupling between $L_X$ and $\dot m$ ensures that the systems remain fixed
in a state as the accretion rate changes. Thus the difference between $L_R/L_X$
correlations is that NS are in a X-ray dominated state and BHs are in a jet dominated
state.

It is generally accepted that very high magnetic fields at the surface of the NS inhibit
the production of -steady- jets  while a large amount of energy can be extracted from
magnetic fields to power extremely energetic -transient- outflows. Although it is not
clear the role of the magnetic field at the surface of a NS in the production of jets,
it is believed that the higher the magnetic field is the lower the jet power should be.

Magnetic fields produced in NS extend from above 10$^{12}$G, in classical X-ray pulsars,
where jets are excluded at any accretion rate \footnote{Different from
millisecond X-ray pulsars where it is not excluded that could develop jets,
at least for sources with $B \leq
10^{7.5}$G \cite{massi-kaufman 2008}. In this case the source could switch to a
microquasar phase during maximum accretion rate. The millisecond source SAX J1808.4-3658
shows indeed hints of a radio jet \cite{migliari-fender 2006}.} \cite{fender-hendry
2000,migliari-fender 2006}, down to 10$^{7-8}$G in other sources connected to XRBs with
jets such as atoll and Z-type systems \cite{massi-kaufman 2008}. In the presence of
jets, the interval of accretion rates also covers a wide range, from less than 0.1\% of
the Eddington  rate for millisecond X-ray pulsars, up to the Eddington critical rate for
Z-sources.

The conditions for an XRB to undergo a microquasar phase, i.e. the generation of jets,
are given by a relationship between the magnetic pressure $P_B=B^2/8\pi$ and the
hydrodynamic pressure $P_p=\rho v^2$, where $\rho$ is the matter density and $v$  its
velocity. First, the magnetic field at the surface has to be relatively weak. Second,
the magnetic field lines have to be already twisted close to the CO \cite{meier al
2001}.
The appropriate regime for jet launching has thus been determined by studying the
quotient between the CO radius and its Alfvén radius $R_A$ (the distance where
$P_B=P_p$). The basic condition for jet formation is  $R_A/R_*$ = 1 ($\ R_A/R_{LSO}$ =
1),  $R_*$ being the neutron star radius and $R_{LSO}$ the last stable orbit of the BH.
This condition implies that the accretion disc arrives down to the surface of the CO,
and hence guarantees that $P_B<P_p$ is valid in the whole disc. It allows quantifying
the upper limit of the magnetic field strength as a function of the mass accretion rate
\cite{massi-kaufman 2008}. As a result of this analysis, for a Kerr BH accreting at the
Eddington rate, the magnetic field cannot exceed $5 \times 10^{8}$G, while in the case
of Z-sources at the same rate, a magnetic field below $B = 10^{8.2}$ G should be in
order to make possible the generation of jets. This upper limit fits the observational
estimative of \cite{titarchuk al 2001} for Scorpius X-1. Atoll-sources, on the other
hand, are potential sources for generating jets provided $B \leq 10^{7.7}$ G. Evidence
of jets in these kind of sources has been already found in \cite{migliari al 2006,
russel al 2007}.

\subsection{VHE gamma-rays emitters}

Some of the VHE gamma-ray sources detected in the last few years \cite{aha hess 2005,
aha hess 2006, albert magic 2006, albert magic 2007, acciari veritas 2008}  have been identified with
previously known X-ray binary systems \cite{hartman al 1999}.
Recently, Cygnus X-3, another well-known microquasar, has been also associated with a
steady high energy gamma-ray source \cite{giuliani al 2008}.  These detections, together
with previous low energy observations, demonstrate the diversity of MHD processes
underlying manifestly correlated non-thermal emission from XRBs with relativistic
outflows.

As we have already mentioned, in a microquasar scenario VHE gamma-rays are supposed to
be produced in the intersection zone common to intra-jet relativistic particles and
external thermal matter plus radiation fields (e.g. \cite{hugo al 2005}), and even fully
within the jet's bulk  \cite{nuestro mnras 2008}. The variability of the emerging signals and
the spectral properties of the sources depend on several conditions and parameters. To
estimate the TeV emission from these systems one has to take into account their
intrinsic possibility of accelerating particles up to the TeV - PeV energy scale, the
different radiative mechanisms, and the absorption of the produced VHE flux. In some
cases, depending on the orbital phase and the intensity of companion and disc photon
fields, the absorption can be strong enough for signals to lie below the sensitivity of
current detectors \cite{nuestro app 2008}.

Inverse Compton processes should produce significant VHE fluxes when external photon
fields penetrate into the jet and are scattered off by VHE leptons. Among IC boosting,
there are contributions to IC from leptonic self-generated synchrotron photons and
relativistic Bremsstrahlung that should enhance the total VHE leptonic output. On the
other hand, when $\rho_{\rm m}>\rho_{\rm k}$, strong synchrotron losses can attenuate this channel
 and reduce gamma-ray production. Intense VHE gamma-ray fluxes  are generated
through the decay of neutral pions from photomeson reactions and from pion production in
relativistic collisions of protons in the jet against cold ions from the wind or the jet
itself.

As explained in  \cite{nuestro mnras 2008}, the gamma-ray luminosity also depends on the
fraction $q_{\rm rel}$ of the kinetic luminosity $L_{\rm k}$ transferred to the accelerated
(relativistic) plasma. There, the authors have estimated the gamma-ray production in the
heavy jets of SS 433 out of relativistic hadronic processes. By using the HEGRA upper
limit on the VHE flux from this source, the value of $q_{\rm rel}$ for accelerated
ultra-relativistic protons is maximally constrained by $q_{\rm rel} < 3\times 10^{-4}$
(considering that HEGRA observations took place during unknown precessional phases). The
last MAGIC observations, performed during the lowest absorption phases allow a more
stringent constraint for the hadronic fraction ($q_{\rm rel}<7.4 ~
10^{-5}$) \cite{saito magic 2009} which, given the expected sensitivity of Icecube, is
still above the lowest testable limit of our model \cite{nuestro mnras 2008}.

\subsubsection{VHE Flares}

Besides steady emission, short-lived ejections are frequent in microquasars. These are
mostly seen in the transitions between the LHS to the HSS, although the flaring
behavior is not limited to the state changes and is quite unpredictable (c.f. GRS 1915
+105 or Cyg X-3). The kinetic power released in such episodes can exceed 10$^{39}$ erg
/s, with intense radiation fluxes at all wavelengths. However, clear signals of VHE
emission from GRS 1915 +105, Cyg X-1, Cyg X-3 and SS 433 remain undetected. In fact,
apart from the steady TeV emission observed in the 4 galactic HMXBs, PSR B1259-63, LS
I+61 303, LS 5039, Cyg X-1, clear evidence of a TeV flare has just been found in Cyg
X-1.   In the case of LS I +61 303, in addition to the periodic TeV emission, with a
maximum at phase 0.6, it has been seen a flaring activity peaking around phase 0.8
\cite{albert magic 2009}. Additionally, there is also a temporal coincidence between the
TeV and the X-ray flare \cite{esposito al 2007} as noted in \cite{paredes heidelberg
2008}. LS 5039 presents also flaring TeV emission superposed to the periodic regular
light-curve around phase 0.8.

The correlation between hard X-ray and TeV emission should provide more close
constraints, since both signals could be produced by the same particle population or
with a similar timescale \cite{saito magic 2009}.
However, the multifrequency correlation could be present only
at some stages of the flare \cite{albert magic 2007}, rendering the start of TeV
outburst detections a difficult task. The non-detection of any transient signal could be
related to a high absorption in the inner regions of these systems. The gamma-ray
attenuation due to stellar and accretion disc photon fields could in any case transfer
the luminosity to lower gamma-ray energies through electromagnetic cascades
\cite{orellana al 2007, bednarek-giova 2007}. This would increase the fluxes at appropriate
ranges for observatories like Fermi, providing information
about the triggering of VHE gamma-ray and its relation with HE radiation in these systems
\cite{abdo fermi 2009}.
It is worth noting also that the detection of
outbursts at radio and IR wavelengths can be related to an increase of the activity also
at VHE, although the delay between them could range from hours to days \cite{atoyan-aha
1999}.


Regarding the variability of the emission, several factors are relevant. Injection can
change due to variations in the accelerator, e.g. injection power and injection spectrum.
Target densities could vary as a consequence of stochastic changes in the magnetic, photon, and
matter fields. Geometry can evolve because of orbital motion or jet precession. Everything
could affect anisotropic gamma-gamma absorption and scattering or photon boosting. Thus,
the time scales of the variability can be linked to a lot of mechanisms including
injection, radiative cooling, particle escape, and macroscopic motion.

So far, it has been experimentally established that the variability of gamma-ray sources
is modulated with the orbital period although short-timescale flares are also observed
\cite{albert magic 2007, paredes ijmp 2008}. As mentioned, the presence of jets in some
of the massive gamma-ray binaries implies that scattering of relativistic particles
against the stellar wind of the primary seems unavoidable \cite{romero al 2003}. Since
there are increasing reasons to think that the winds of hot stars have a clumped
structure (e.g. \cite{dessart-owocki 2003, puls al 2006}) if the jet interacts with the
stellar wind, the putative gamma-ray emission would present a variability related to the
structure of the wind. Thus, the detection of rapid stochastic variability, quite
distinct from long-term periodic variations, as orbital and precessional, could be used
to understand the structure of the wind itself \cite{owocki al 2009}.

Notwithstanding, intrinsic disturbances in the jet (see e.g. \cite{mirabel al 1998,
marscher al 2002})  can also produce an aperiodic variability that might be confused with
jet/clump interactions. Fortunately, as signaled in \cite{owocki al 2009}, intrinsic
variability in the jet would likely be preceded by a change in the accretion disc X-ray
activity, whereas in the case of a jet/clump interaction, the effect should be the
opposite; first, the gamma-ray flare would appear, and then, a nonthermal X-ray flare
produced by the secondary electrons and positrons (as well as the primary electrons
injected into the clump) would show up. Depending on the magnetic field and the clump
density, the X-ray radiation could be dominated by synchrotron, inverse-Compton, or
Bremsstrahlung emission, with a total luminosity related to that of the gamma-ray flare.
Simultaneous X-ray observations with gamma-ray observations could be useful to
distinguish jet/clump events from intrinsic variability produced by the propagation of
shocks in the jets.

\subsection{Neutrinos}

In general, arguments supporting predictions of discrete neutrino sources are based on
the so-called $\gamma$-$\nu$ connection \cite{dermer 2006}. Photohadronic processes and
proton scattering off nuclei are the most important astrophysical neutrino production
mechanisms and they also produce gamma-rays. By identifying the brightest steady
gamma-ray sources, we identify the most likely neutrino point sources to be surveyed. Of
course, a source can be gamma-ray intense without being neutrino significant if the
gamma-rays originate from leptonic processes. Conversely, a bright neutrino source can
be gamma-ray dim if photon production is attenuated at the origin or along its path. At
EGRET and GLAST energies ($E \sim$ 100 MeV - 10 GeV), attenuation by the extragalactic
background is unimportant, even for the highest redshift objects. Thus, the EGRET
catalog should be exhausted to identify the brightest gamma-ray sources and, by the
above argument, the most probable neutrino point sources. The new discoveries
with HESS, MAGIC, and VERITAS at gamma-ray TeV energies are especially significant for
understanding and searching other galactic neutrino sources.

High-energy neutrino production takes place in a hadronic scenario
and results from proton-ion interactions and photohadronic processes.
The dominant channel for neutrino production involves the production of pions coming
from these reactions. On average, neutron plus charged-pion reactions take place one
third of the time while proton plus neutral-pion production occurs on two thirds. The
outcome is therefore one high-energy lepton and three high-energy neutrinos for every four
high-energy gamma rays  (neutron decay produces a neutrino but just $\sim 1\%$ energetic,
in favor of the resulting proton momentum \cite{dermer 2006}).
Since neutrinos carry only 1/4 of the energy radiated in
electromagnetic secondaries, any high-energy neutrino source should be a strong gamma-ray
source, were it not for gamma-ray opacity which can seriously modify this expectation
\cite{nuestro app 2008}.

For such gamma-ray sources, the selected neutrino sources should be those which IceCube
could discriminate above the cosmic-ray neutrino induced background. At $E\geq$1 TeV,
this background is at $n_\nu\geq$ 0.4 neutrinos/yr per square degree, and twice
this value for $E\geq$10 TeV \cite{karle al 2003}. To be
detected with a km-scale facility, the neutrino flux, and therefore the photon flux,
must be above $10^{-4}$ ergs/cm$^2$ \cite{dermer-atoyan 2006}.

In the Third EGRET catalog \cite{hartman al 1999}, several TeV gamma-ray sources that fit
this criterion can be found, particularly 3EG J1824-1514 \cite{paredes al 2000} and 3EG
J0241+6103, respectively associated with the HMXBs LS 5039 \cite{aha hess 2005} and LSI +61 303
\cite{albert magic 2006}. Models for these gamma-ray emitters as potential neutrino
sources have been developed in
\cite{nuestro prd 2006, torres-halzen 2007, bosch 2007, neronov-ribordy 2009} and
their microquasar nature inspired also neutrino predictions in systems like SS 433
\cite{nuestro mnras 2008} though not found yet in the TeV range, probably due
to intense local opacity \cite{nuestro app 2008}.

Present and upcoming experiments like ICECUBE (e.g. \cite{abassi al 2009}),
ANTARES (e.g. \cite{van elewyck 2009}) and NEMO (e.g. \cite{distefano al 2007}) are expected
to show exciting new horizons in neutrino astronomy.



\section{model for high energy emission in XRBs with jets}\label{model}

In order to make a model approach to high energy emission in binary systems with jets,
some geometrical assumptions are in order. Following \cite{hjellming-john 1988} we can assume
that jets are well collimated and conical-shaped, with small half-opening angles
($\xi\leq 5^o$). The jet radius thus reads $r(z)= z\tan\xi $ and its structure is
completely defined all along. It is also reasonable to assume the jet injection point at
a distance $z_I = 50 R_g$  where
$R_g=GM_{C}/c^2$ is the gravitational radius of the ejecting object
(typically $\simeq 10^{7-8}\rm cm$ from a stellar CO).


The jet power and the accretion rate are related by $\dot M_j=\frac 1 2 q_j\ \dot m$,
assuming $\dot m=q_{\rm a}\ L_{{\rm Edd}}$ and $q_{j}, q_{\rm a} < 1$  (the 1/2 factor
accounts for the
existence of a counter-jet). The kinetic luminosity of the jet can be written as $ L_{J\rm
k}={\dot M_{j}/ E_{\rm b}}{m_{p}}$ where $E_{\rm b}$ is the bulk kinetic energy and $m_p$ the
 jet particle mass. The jet kinetic energy density is $\rho_{\rm k}(z)= {L_{J\rm k}}/{\pi
r_{j}^2 v_{\rm b}},\ \rm(erg/cm^3)$ where $v_{\rm b}$ is the velocity
(the subindex ${}_{\rm b}$ is for bulk quantities).
Following with the jet-accretion coupling hypothesis, we assume that a considerable part
of the Eddington luminosity ($\sim$ 10\%) goes into the jet \cite{kording al 2006} so
that $L_{J\rm k}$ could attain up to $\sim 10^{38}{\rm erg \
s}^{-1}$. This power is carried among leptons and hadrons, a small fraction of which
will be ultra-relativistically accelerated within the jet.

As we have seen, the jet can be divided in several sectors where particles are
accelerated and cooled by different means. Following \cite{protheroe 1999} the
stochastic motion in a collimated outflow results in a mean electric field in the $z$ direction.
Defining the rate of energy variation by means of $t^{-1}=E^{-1}dE/dt,$ the rate of
stochastic electromagnetic acceleration of a $eZ$-charged particle up to an energy $E$
is given by $t_{\rm acc}^{-1} \approx \eta {c \; eZ \; B}/{E_p}\label{tacc}$ (CGS),
where $\eta$ is the efficiency of the accelerator (e.g. \cite{begelman al 1990}). In the
energy equipartition region, where the magnetic energy density $\rho_{\rm m}$ equals the
kinetic density $\rho_{\rm k}$, the magnetic field reads $B(z_E)= \sqrt{8\pi \rho_{\rm
k}(z_E)}= {2\sqrt{\dot M_{j} v_{\rm b}}}/ {z_E\tan\xi}$ and reaches very high
values. It varies with distance as
$(z_E/z)^m, \ 1\leq m\leq 2$ \cite{krolik 1999}. Here we assume that just above the jet
injection point, $z_I$, the jet is magnetically dominated while in the
acceleration/transport zone, $z_0\leq z\leq\sim 5z_0$ ($z_0 \sim 10 z_I$), the magnetic
field is weaker and the jet is dynamically dominated, $\rho_k >\rho_m$ (see
\cite{komissarov al 2007}). In this zone we have to consider a transport equation in
order to take into account energy losses as well.

Assuming shock diffusive acceleration near the base, the injection rate is a canonical
second-order power-law in the particle energy. On the other hand, conservation of the
number of particles demands that the current evolves with $z^2$. Hence,
$J'(E',z')=cK_0/4\pi\ E'^{-2} (z_0/z)^2$ \cite{ghisellini al 1985}, where primed
quantities are in the jet frame. In the absence of sinks or sources, other than the
termination and injection points, the injection function of particles,
$Q(E,z)$, must satisfy a continuity relation with the current \cite{magnetico 2009}.
After some relativistic considerations,  for this function we obtain
\bea Q(E,z)= R_{0}\left(\frac{z_0}{z}\right)^3\frac{\Gamma_{\rm
b}^{-2}\left(E-\beta_{\rm b} \sqrt{E^2-m^2c^4} \cos {i_j} \right)^{-2}}{\sqrt{\sin ^2
{i_j} + \Gamma_{\rm b}^2 \left( \cos {i_j} - \frac{\beta_{\rm b} E}{\sqrt{E^2-m^2
c^4}}\right)^2}}\label{peinjection} \eea
in an observer's reference frame making an angle $i_j$ with
the line of sight\footnote{Note that the angle $i_j$, as well as $Q$, can
depend on time. In the first case this is easy to handle since it is a function of
precession; in the second it depends on several factors, including instabilities and
stochastic processes.}. The normalization constant $R_{0}$ is the
power density at the acceleration/injection point $z_0$  and results by
specifying the relativistic lepton (proton) luminosity of the jet \be
  L_{e,p}= \int_V d^3r \ d\Omega\
  \int_{E_{e,p}^{\rm(min)}}^{E_{e,p}^{\rm(max)}}dE_{e,p}
  E_{e,p}\ Q_{e,p}(E_{e,p},z),\label{qrelNorm}
\ee where $E_e^{\rm(min)}= 1 $ MeV, $E_p^{\rm(min)}=1.2$ GeV, and the maximum energies
will be obtained by equating the acceleration rate to the energy loss rate.
The fraction of the total kinetic power carried by the ultra-relativistic primary
particles in the jet is parameterized by means of $q_{\rm rel}$, namely $q_{\rm rel} L_{\rm k}
= L_{\rm rel}= L_p + L_e$. We can further assume a simple relation between the proton
and electron luminosities as given by an hadronicity parameter, $h$, such that $L_p= h \
L_e$ \cite{romero-vila 2008}.

\subsubsection{Energy losses}

The main particle processes responsible for the energy cooling in a leptonic/hadronic
jet are the following.

\vskip 0.2cm
\textit{Synchrotron cooling}: $ e^{\mp} + (B)\rightarrow e^{{\mp}'}  + \gamma'
\label{electronsynchro}$ \vskip 0.2cm Of course, fast moving charged particles emit
strong synchrotron radiation\footnote{Synchrotron radiation is an analog to
bremsstrahlung, differing in that the force which accelerates the electron is a
macroscopic or large scale magnetic field. Bremsstrahlung is the electromagnetic
radiation produced by a sudden slow down or deflection of charged particles, especially
electrons, passing through matter in the vicinity of the strong electric fields of
atomic nuclei or ions. In a broad sense, bremsstrahlung is the radiation emitted when
any charged particle is accelerated by any force, but to a great extent, is a source of
photons in the ultraviolet and soft x-ray region for the investigation of atomic
structure.}. Its rate is given by $t_{\rm
sync}^{-1}=\frac{4}{3}\left(\frac{m_e}{m}\right)^3 {\sigma_{\rm T}B^2\gamma }/{m_e c \
8\pi}\label{tsyn}$ (e.g., \cite{begelman al 1990}),
where $m$ is the mass of the particle and $E=\gamma \ m c^2$ its
relativistic energy.

The power per unit energy of the synchrotron photons radiated by  $N_{e,p}(E,z)$ charged
particles is $\varepsilon_{\rm syn}^{(e,p)}(E_\gamma)= \int d\Omega_\alpha
  \int_{E_{e,p}^{\rm(min)}}^{E_{e,p}^{\rm(max)}}dE_{e,p}\ P_{\rm syn}
  N_{e,p}(E,z),$ where $P_{\rm syn}(E_\gamma,E,z,\alpha)$ is the power radiated
by a single particle of energy $E$ and pitch angle $\alpha$ (e.g. \cite{blumenthal-gould
1970}) and the total luminosity can be obtained by integrating in the volume of the
region of acceleration $
 L_{\rm syn}^{(e,p)}(E_\gamma)= \int_V d^3 r \; E_\gamma\ \varepsilon_{\rm
syn}^{(e,p)}.
$

\vskip 0.5cm
\textit{Inverse-Compton cooling}: $ e^{\mp} + \gamma \rightarrow e^{{\mp}'}
+ \gamma'  \label{inversecompton} $
 \vskip 0.2cm
In turn, ambient photons serve as targets for the electrons themselves and the
corresponding rate is given by $t_{\rm IC}^{-1}(E,
z)=\frac{4}{3}\left(\frac{m_e}{m}\right)^3 {\sigma_{\rm T}\rho_{\rm ph}\gamma}/{m_e c},$
where $\rho_{\rm ph}=\int d\epsilon\ \epsilon \;{N_{\rm ph}(\epsilon, z)}$, integrated
in ${\epsilon<{m_p^2c^4}/{E}}$, is the photon energy density \cite{begelman al 1990}. In
the case of self-synchrotron Compton scattering, the radiation density can be
approximated by $N_{\rm ph}(\epsilon, z)\approx n_{\rm{synchr}}(\epsilon,z)=
\frac{\varepsilon_{\rm syn}}{\epsilon} \frac{r_{\rm j}(z)}{c},$ as in \cite{ghisellini
al 1985}.

\begin{figure*}
\includegraphics[trim = 0mm 0mm 0mm 0mm, clip,height=.5\textheight,
width=0.8\linewidth,angle=0]{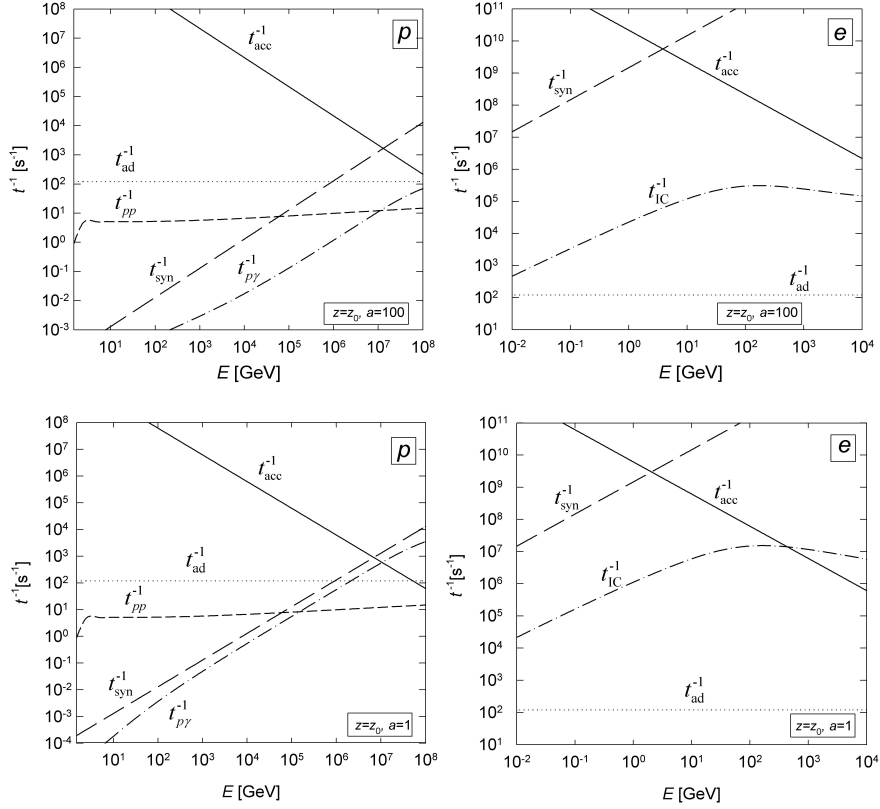}
  \vspace{-10mm}
\caption{Accelerating and cooling rates for protons (left) and electrons (right) at the
base of a hadro-leptonic jet dominated by protons ($h=100$, top) and with power
equipartition ($h=1$, bottom). Figures from \cite{magnetico 2009} ($a$ is the
hadronicity $h$ in the present notation).} \label{fig rate p-e}
\end{figure*}

The IC spectral luminosity reads
$$L_{\rm{IC}}\left(E_\gamma\right)=E^{2}_\gamma \int_V
\mathrm{d}^3r\int_{E_{}^{\rm{min}}}^{E_{{e}}^{{max}}
\left(z\right)}\mathrm{d}E_{{e}}\,N_{e,p} \int_{\epsilon_{{min}}}^{\epsilon_{{max}}}
\mathrm{d}\epsilon\,P_{\rm{IC}}, \label{ICluminosity} $$
where $P_{\rm{IC}}\left(E_\gamma, E, \epsilon, z\right)$ is the spectrum of photons
scattered by a charged particle of energy $E$ \cite{blumenthal-gould 1970}.

\vskip 0.3cm
Regarding protons, synchrotron and pure IC loss rates are more than 9 orders of magnitude less
significant than for electrons, and the principal energy cooling mechanisms are the following.

\vskip 0.5cm
\textit{Photopion production}: $ p+\gamma\rightarrow p+a\pi^0+b\left(\pi^++\pi^-\right)
\label{photomeson1}$

\hskip 3.9cm   $p+\gamma\rightarrow n+\pi^+ + a'\pi^0+b'\left(\pi^++\pi^-\right),
    \label{photomeson2}$

\vskip 0.2cm
\noindent where $a, b, a', b'$ are the pion multiplicities of each channel \footnote{At
lower energies, one could also consider direct IC photon boosting by protons $p + \gamma
\rightarrow p'+ \gamma' \label{p inversecompton}$ (see \cite{atoyan-dermer 2003}) but it
is irrelevant at high energies even in extreme radiation fields.}.

The photopion cooling rate is given by
$$ t_{p\gamma}^{-1}(E,z)=\frac{c}{2\gamma_p}\int_{\frac{e_{\rm th}}{2\gamma_p}}^{\infty}
d\epsilon\frac{N_{\rm ph}(\epsilon,z)}{E_{\rm ph}^2} \int_{\epsilon_{\rm
th}}^{2\epsilon\gamma_p} d \epsilon'
  \sigma_{p\gamma}^{(\pi)}(\epsilon')K_{p\gamma}^{(\pi)}(\epsilon')
  \; \epsilon' \label{tpg}
$$ where, $\epsilon_{\rm th}=145$ MeV \cite{begelman al 1990}.
Approximated expressions for the cross section $\sigma_{p\gamma}^{(\pi)}$ and
inelasticity can be found in \cite{atoyan-dermer 2003}.
Photopion production is the predominant $p\gamma$ channel. It mainly occurs when protons
collide with X-ray photons from the corona ($2{\rm keV}< E < 100$keV), for which, based
on \cite{cherepashchuk al 2005}, we can adopt a Bremsstrahlung X-ray distribution
$n_{\rm X}(E)=L_{\rm X} {e^{-{E}/{(kT_e)}}}/{4\pi z_{\rm j}^2 E^{2}}
 ({\rm erg}^{-1}{\rm cm}^{-3}),\label{Xrays}$
where $kT_e \approx 30$keV and $L_X=10^{36}{\rm erg}/{\rm s}$.

\vskip 0.5cm
\textit{Photopair production}: $p+\gamma\rightarrow p+e^-+e^+,\label{p photopair}$ is
calculated using the same $t_{p\gamma}^{-1}$ expression, but the soft photon density
includes also the contribution of UV emission (particularly if there is an extended
disc, as in SS433), $n_{\rm ph}(E)= {2E^2}{(h c)^{-3}( e^{E/kT_{\rm UV}}- 1)^{-1}}{{\pi
R_{\rm out}^2}/{z^2}}+ n_{\rm X}(E)$, with $T_{\rm UV}=21000$ K, based on \cite{gies al
2002}. For this process we consider the expressions for cross section
$\sigma_{p\gamma}^{(e)}$ and inelasticity $K_{p\gamma}^{(e)}$ given in \cite{begelman al
1990}. See also \cite{kelner-aha 2008} for a detailed discussion.

\begin{figure*}
\includegraphics[trim = 0mm 0mm 0mm 0mm,
width= 0.5\linewidth,angle=0]{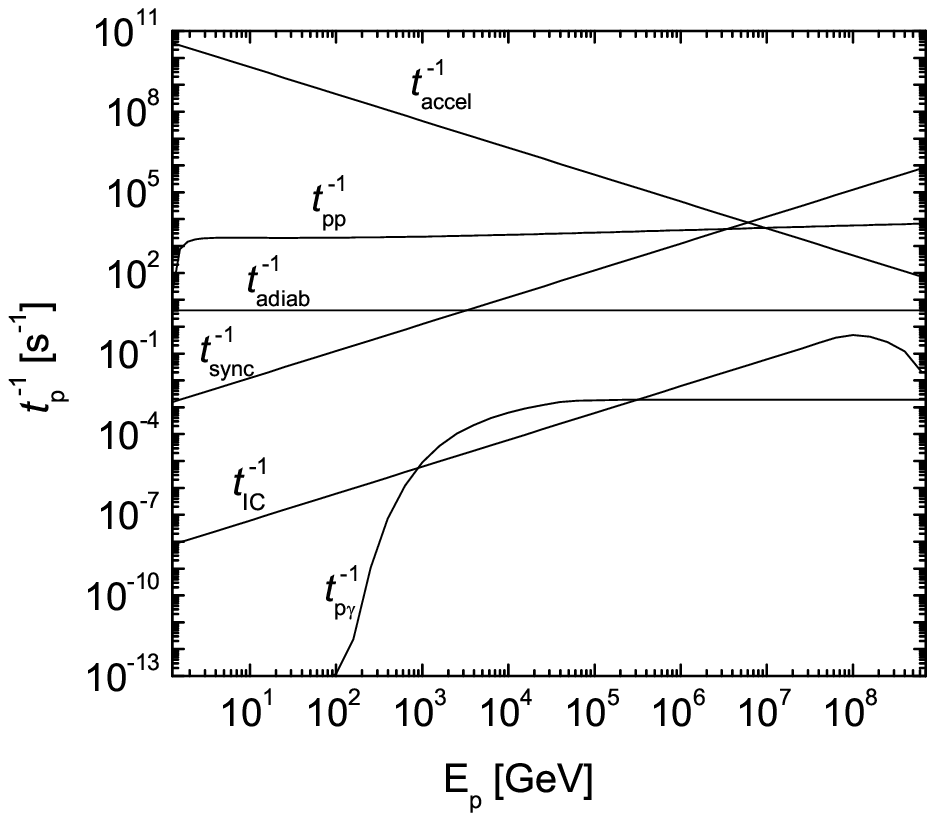}
\includegraphics[trim = 0mm 0mm 0mm 0mm,width= 0.5\linewidth,angle=0]{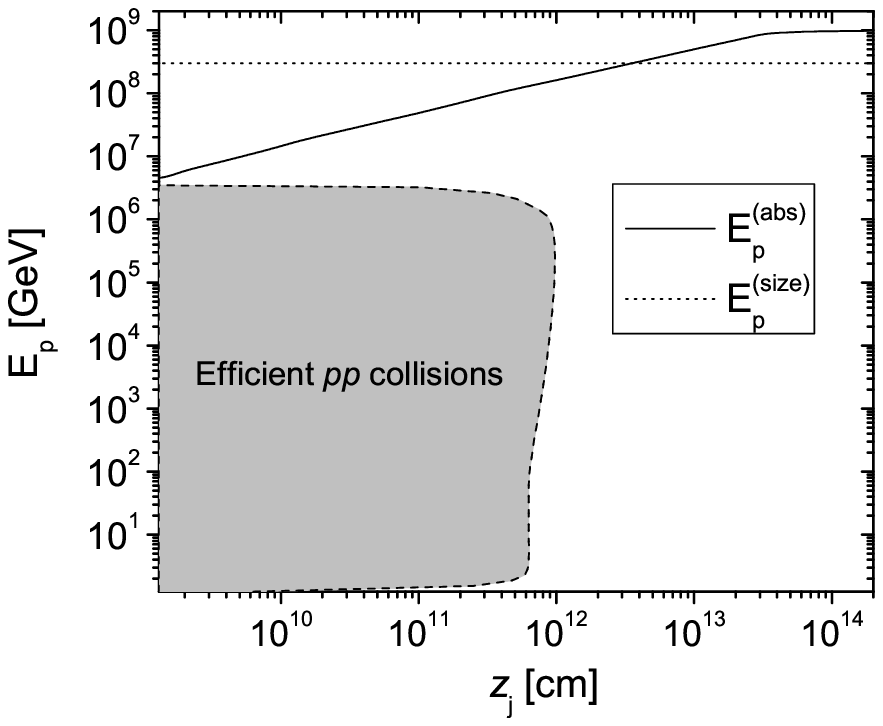}
\caption{(Left) Proton accelerating and cooling rates. (Right) $E$-$z$ region where pp
collisions are dominant. Note maximal proton energy values $E_p^{\rm abs}$ obtained from $t_{\rm
acc}^{-1}= \sum t_{\rm loss}^{-1}$. Details in \cite{nuestro mnras 2008}.} \label{fig
rate pp}
\end{figure*}

\vskip 0.5cm
{\textit{Proton-proton cooling}}:
$p+p\rightarrow p+p+A\pi^0+B\left(\pi^++\pi^-\right)
    \label{pp1}$

\hskip 4cm $ p+p\rightarrow p+n+\pi^+ +A'\pi^0+B'\left(\pi^++\pi^-\right),
    \label{pp2}$

\hskip 4cm $ p+p\rightarrow n+n+2\pi^+ +A''\pi^0+B''\left(\pi^++\pi^-\right),
    \label{pp3}$

\vskip 0.2cm
\noindent where $A, B, A', B', A'', B''$ are the pion multiplicities of each channel.
Pure hadron collisions are crucial, particularly for VHE production. Now, the rate of
$pp$ collisions of ultra-relativistic protons against the cold ones is given by
$t_{pp}^{-1}= n(z) \; c \; \sigma_{pp}^{\rm(inel)}(E_p)K_{pp}, $ where the inelasticity
is $K_{pp}\approx 0.5$, assuming that the projectile yields half of its energy for
secondaries. The density of cold target particles in the jet at a distance $z\geq z_0$
from the compact object is  $n(z)= {(1-q_{\rm rel})L_{\rm k}}/{\Gamma_{\rm b} m_p c^2
\pi r_{\rm j}^2 v_{\rm b}}$  and the cross section ${\sigma_{pp}^{\rm(inel)}} $ is given
in \cite{kelner al 2006}.

\vskip 0.3cm
Finally, since the jet is laterally expanding with velocity $v_{\rm b}\tan \xi$, there
is also an \textit{{adiabatic cooling}} rate given by $ t_{\rm ad}^{-1}(z)=
\frac{2}{3} {v_{\rm b}}/{z}$ \cite{bosch al 2006}.

The  highest energies theoretically reached by the particles can be obtained
from an equation $ t_{\rm acc}^{-1}(E^{\rm(max)})= \sum t_{\rm
loss}^{-1}(E^{\rm(max)})$ for each species of particle.
Depending on the position $z$ and the hadronicity $h$, one obtains different values
of $E^{\rm(max)}$ for leptons and for hadrons. See the results in
Fig. \ref{fig rate p-e} and Fig.\ref{fig rate pp}.

\begin{figure*}
\includegraphics[trim = 0mm 0mm 0mm 0mm, clip,height=.3\textheight,width=.6\linewidth,
angle=0]{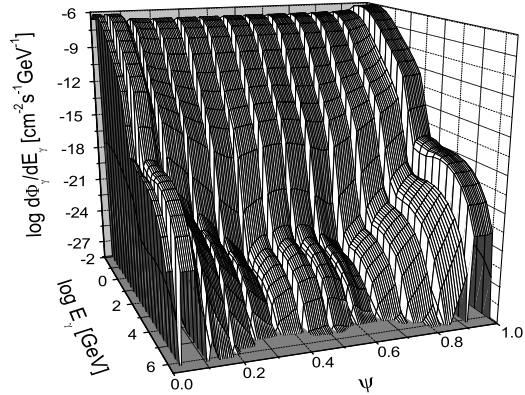}
  \vspace{-10mm}
\caption{Differential gamma-ray flux from microquasar SS 433 arriving at Earth as a function of
precessional phase and energy. In this case, (strong) absorption effects have
been included. See details and discussion in
Ref.\cite{nuestro mnras 2008}.}
\label{MN-08-0074-MJ-Fig9}
\end{figure*}

\begin{figure}
\includegraphics[trim = 0mm 6mm 0mm 8mm, clip, height=.25\textheight,
width=0.5\linewidth,angle=0]{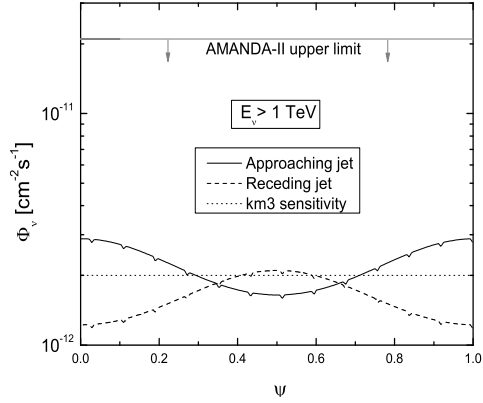}
\vspace{-10mm}
\caption{Neutrino flux ($E_\nu>1$ TeV) arriving at Earth as a
function of the jet precessional phase. The contributions of hadronic
jet and counterjet are shown separately, solid line for the approaching jet and
dashed line for the receding one (secondary losses neglected). The upper limit from AMANDA-II
data and the expected km$^3$ sensitivity are shown in grey solid
line and dotted line respectively.
Figure from \cite{nuestro mnras 2008}}\label{Fignut}
\end{figure}

\subsubsection{The transport equation for primaries}

In each acceleration sector of the jet, and for each particle species,
the steady state spectral primary particle
distribution can be obtained as the solution of a transport equation
 \be
 \frac{\partial{[N_i(E,z)b_i(E,z)]}}{\partial E}+ \frac{N_i(E,z)}{T_{i}^{\rm esc}(z)}=
 Q_i(E,z), \label{teq}
 \ee
where $b_i(E,z) \sim  -E \sum t^{-1}_{i\ \rm loss}(E,z)$ and
$ T_{i}^{\rm esc}(z)\approx ({z_{i}^{\rm max}-z})/{c} $ is the escape rate from sector
$i$ ($z_{i}^{\rm max}$ is the termination point of this sector). The corresponding
solution is
$$N_i(E,z)= \frac{1}{|b_i(E)|}\int_E^{E_i^{(\rm max)}}dE'Q_i(E',z)
\times \exp{\left\{-{\tau_i(E,E')}/{T_{i}^{\rm esc}}\right\}} \label{Nep} $$
with $ \tau_i(E,E')= \int_E^{E'} {dE''}/{|b_i(E'')|}$ (e.g. \cite{khangu al 2007}).
The effect of particle
acceleration is implicit in the (ultra-relativistic) injection function $Q_i$
 and the effect of deceleration
is in the $b_i$ function. As a matter of fact, a complete description of the evolution
requires a time dependence of the functions, but a solution is still lacking. In each
sector $i$ of the jet and for each kind of particle the dominant acceleration mechanisms
and cooling processes vary, as we have already explained. In order to simplify
calculations though extracting enough dynamical features, we can assume a one zone
acceleration approach and consider a small sector near the base $z\geq\sim z_0$, where
shocks are more frequent. Taking into account a lower acceleration efficiency, a similar
approximation can be exploited in the case of a clumpy wind (typically in HMXBs)
colliding with the jet at middle scales (see \cite{araudo al 2009}).

\subsubsection{Secondary particles and VHE luminosity}
As we have mentioned above, the primary relativistic protons produce
pions through inelastic interactions with matter and radiation. Charged pions in turn
decay into muons and neutrinos, and muons then go into neutrinos and electrons:
$$ \pi^- \rightarrow \mu^-\bar{\nu}_\mu \rightarrow e^-\nu_\mu \bar{\nu}_e
  \bar{\nu}_\mu, \ \
  \pi^+ \rightarrow \mu^+ {\nu}_\mu \rightarrow e^+\bar{\nu}_\mu \nu_e
  {\nu}_\mu.
$$
On the same footing, neutral pions decay into gamma-rays through the channel $$ \pi^0
\rightarrow \gamma\gamma.$$

The distribution of energetic gamma-rays results from the distribution of neutral pions. As a
result of $pp$ interactions, the gamma-ray emissivity, in units ${\rm GeV}^{-1}{\rm
s}^{-1}$, reads
$$ {dN_\gamma(t, E_\gamma, z)}/{dE_\gamma}= \int_{x_{\rm min}}^{x_{\rm
max}} \sigma_{pp}^{\rm inel}\left(\frac{E_\gamma}{x}\right) J_p\left(t,
\frac{E_\gamma}{x}, z\right) F_\gamma\left(x,\frac{E_\gamma}{x}\right)  dx$$
where the
spectrum of produced gamma-rays $F_\gamma(x,E_p)$ is given in \cite{kelner al 2006},
based on SIBYLL simulations of $pp$ interactions including perturbative QCD effects; $x$
is defined by $E_\gamma= x E_{p}$, for a primary proton with energy $E_p$, and the
integration limits are chosen in order to cover the proton energy range where $pp$
collisions dominate at each $z$.
The spectral intensity of gamma-rays emitted from the jet can be
obtained from
$$
I_\gamma(t,E_\gamma)= \int_{z_0}^{z_1}dz_{}\ \pi(z_{}\tan \xi)^2 n_p
{dN_\gamma(t, E_\gamma, z_{})}/{dE_\gamma}.
$$

As for gamma-rays coming from photopion production at very high energies, the gamma-ray
emissivity reads
$$ {dN_\gamma(t, E_\gamma, z)}/{dE_\gamma}= \int N_{p}(E_p)n_{\rm
ph}(\epsilon) \Phi(y, x)d\epsilon\ dE_p/E_p.$$
Here, the distribution of the target
photons of energy $\epsilon$ and the function $\Phi$ can be found in \cite{kelner-aha
2008} based on SOPHIA Montecarlo routines\footnote{As a matter of fact, one should also
include kaon and $\eta$ channels since they contribute noticeably in the production of
secondaries's gamma-rays and leptons. These have accordingly been taken into account in the
simulations \cite{kelner-aha 2008}.};  $y=4\epsilon
E_p/m_p^2c^4 \geq 0.313$.

Naturally,  before decaying, secondaries are also free to interact (at a rate depending
on the ambient density). Charged pions and leptons will be accelerated as
described in the previous section, but they will also  lose energy and emit radiation
according to the processes discussed above. See contributions at different energies
in Fig.\ref{fig vila-romero09}. On the other hand, the main channel for neutrino production
is through charged pions into muons.

In the case of pions, one has $b_\pi(E,z)= -E( t^{-1}_{\rm syn}+t^{-1}_{\pi p}+
t^{-1}_{\pi\gamma}+ t^{-1}_{\rm ad}).$ For the $\pi p$ interactions $t^{-1}_{\pi
p}(E,z)\approx {n(z) \; c \; \sigma_{\pi p}^{\rm(inel)}(E_p)}/{2},$ where $\sigma_{\pi
p}^{\rm inel}(E) \approx ({2}/{3}) \sigma_{pp}^{\rm inel}(E)$ based on the fact that the
proton is composed by three valence quarks while the pion is formed by two \cite{gaisser
1990}. As for the $\pi \gamma$ interactions we estimate a cooling rate using the
expression of $ t^{-1}_{p\gamma}$ properly adapted with the same argument (e.g. by
replacing $\sigma_{p\gamma}^{(\pi)}\rightarrow ({2}/{3})\sigma_{p\gamma}^{(\pi)}$).
Regarding muons,
$ b_\mu(E,z)= -E ( t^{-1}_{\rm syn}+ t^{-1}_{\rm ad}+t^{-1}_{\rm
IC})$, and we can ignore their interactions with primaries or pions.
Since secondaries have short lifetimes, in the corresponding transport equations
we have to consider $t_{\rm esc}=\rm min
\{T_{\rm esc}, T_{\rm dec}\}$ with $T^{-1}_{\rm dec}=[2.6\times 10^{-8}
\gamma_\pi]^{-1}({\rm s}^{-1})$ for pions and $T^{-1}_{\rm dec}=[2.2\times 10^{-6}
\gamma_\mu]^{-1}({\rm s}^{-1})$ for muons.


The injection function for pions produced either by $pp$ or $p\gamma$ interactions,
$Q_\pi^{(pp)}(E,z)$ and $Q_\pi^{(p\gamma)}(E,z)$, can be found in \cite{kelner al 2006}.
The corresponding  pion distributions, $N_\pi^{(pp)}(E,z)$ and
$N_\pi^{(p\gamma)}(E,z)$, obey an independent transport equation with the obvious
replacements. We can use these to compute the different emitting processes for each
charged species, and also as a source for generating the injection distributions of
descendant muons.

\begin{figure*}
\includegraphics[trim = 0mm 0mm 0mm 0mm, clip,height=.27\textheight,width=\linewidth,
angle=0]{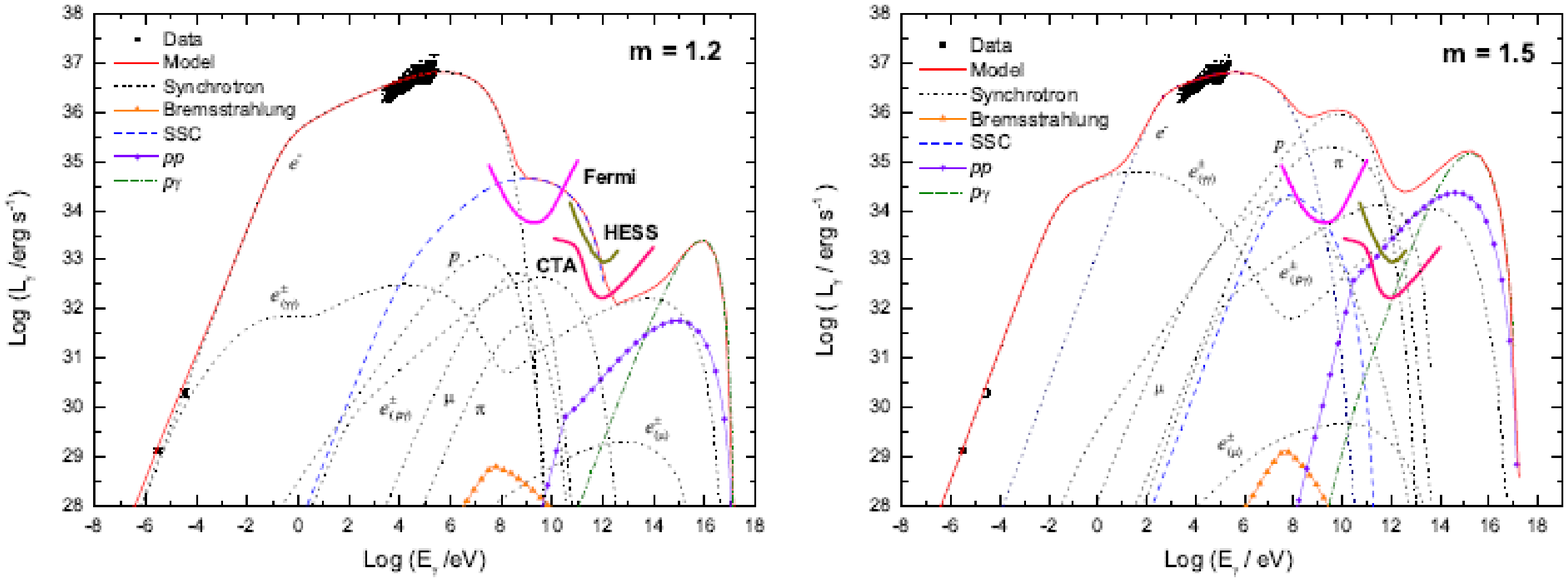}
\vspace{-10mm}
\caption{Best-fit spectral photon energy distributions for microquasar GX 339-4 ($m=$ is
the magnetic field decay index). The subindexes ($\gamma\gamma$), ($p\gamma$) and
($\mu$) indicate pairs created through photon-photon annihilation, photopair production
and muon decay, respectively. The thick lines are the sensitivity limits of Fermi and
HESS, and the predicted for CTA. Figures from \cite{vila-romero 2009}. }\label{fig
vila-romero09}
\end{figure*}

In the case of muons, since right-handed and left-handed species have different decay
spectra, ${dn_{\pi^\pm \rightarrow \mu^\pm_L;\mu^\pm_R}(E_\mu, E_\pi)}/{dE_\mu}$, it is
necessary to consider  their production separately, as discussed in \cite{lipari al
2007}. Each of the resulting injection functions are put in the
corresponding transport equations and, after adding helicities, one obtains the final
spectral distributions of muons of each charge, $N_{\mu^\pm}$, originated from charged
pions of both channels.

The total emissivity is the sum of the contribution of direct pion decays plus that of
muon decays ${dN_\nu(E,z)}/{dE_\nu}= {dN_\nu}/{dE_\nu}^{\pi\rightarrow\nu}+
{dN_\nu}/{dE_\nu}^{\mu\rightarrow\nu}$ (see \cite{lipari al 2007}).
The differential flux of neutrinos arriving at the Earth is
${d\Phi_\nu}/{dE}= \frac{1}{4\pi d^2}I_\nu(E),$
where $d$ is the distance to the source and the neutrino intensity
 $I_\nu(E)= \int_V d^3r\; {dN_\nu(E,z)}/{dE_\nu}$ (${\rm GeV}^{-1}{\rm s}^{-1}$)
 depends on $h$ and the jet half-opening angle $\xi$.

\begin{figure}
\includegraphics[trim = 0mm 0mm 0mm 0mm, clip,height=.3\textheight,
width=\linewidth,angle=0]{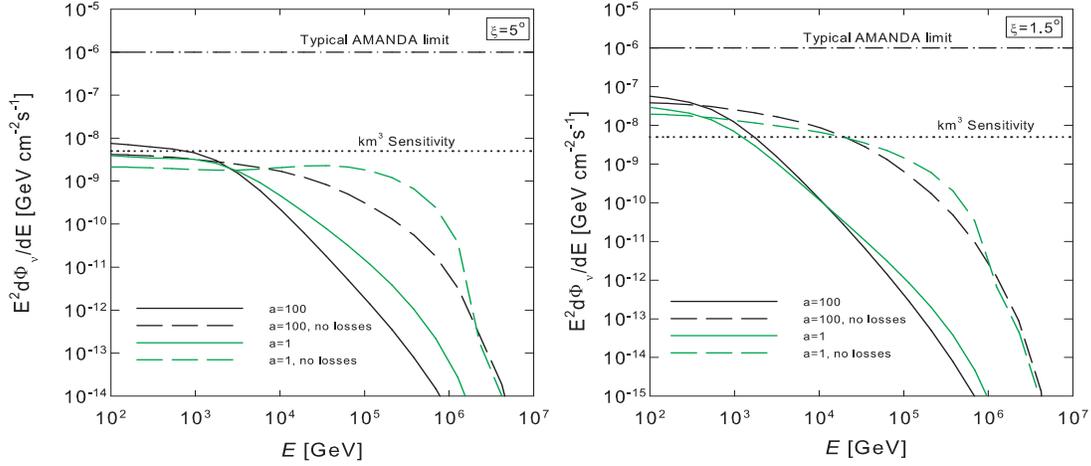}
\vspace{-10mm}
\caption{Differential neutrino fluxes (weighted by energy squared). Cases
$\xi=5^\circ,\ 1.5^\circ$ are shown in the left and right panels respectively.
Black lines correspond to $h=100$ and green lines to $h=1$. Solid (dashed) lines: losses
of secondary pions and muons considered (neglected). Figure from \cite{magnetico 2009}.}
\label{FigE2dflu-base}
\end{figure}

This quantity, weighted by the squared energy, is shown in Fig. \ref{FigE2dflu-base} for
a source at $d=2$ kpc, different values of the jet opening angle, and
different values of $h$. As a guide, we also include a typical upper limit as derived
from AMANDA-II data, as well as the expected sensitivity for the next generation
neutrino telescope (\cite{halzen 2006}, see also \cite{aiello al 2007}).

\subsection{Non-XRB VHE sources}

In addition to X-ray stellar-mass binaries with a compact object, either galactic and
extragalactic star-forming regions or active galactic nuclei and gamma-ray bursters
are strong
candidate sources for very high-energy emission.

\subsubsection{Jets and outflows from young stellar objects}

So far we have been discussing outflows ejected from neutron stars and black holes,
namely dead stars. However, star-forming regions (SFR) can also develop relativistic
 jets and gamma-rays can be produced through different processes along their
evolution, from dark clouds to open clusters \cite{araudo al 2008}.
%
From the experimental point of view, the discovery of the extended source TeV J2032+4130
in Cygnus OB2 by HEGRA in 2002 \cite{aha al yso 2002} marks a cornerstone on this
subject. Recently, two other such gamma-ray sources have been detected, namely,
Westerlund 2 \cite{aha al yso 2007} and W43 \cite{chavez 2008} revealing that SFR are
significant candidates for the new generation of gamma-ray Cherenkov telescopes.

The jets of a massive young stellar object (YSO) propagate through the molecular cloud
where a protostar is embedded, either ending in the interior of the cloud or breaking
out at its surface where strong shocks are expected to occur. Even if obscure at optical
wavelengths, jets have been observed in radio, revealing their characteristic long scale
features (e.g. \cite{marti al 1995}).
In early stages of star formation, the integrated luminosity of many individual
protostars in a massive dark cloud can reach values of 10$^{33}$ erg/s at gamma-ray
energies up to 100 MeV \cite{bena-romero 2003}. Actually, the simultaneous emission of
several such YSOs, can illuminate the cloud in gamma-rays producing a detectable source
for instruments like EGRET and AGILE, while higher resolution telescopes like LAT
(GLAST-Fermi satellite) are expected to resolve even the individual sources.

When massive stars are already formed, the collective effect of many stellar winds is
expected to result in particle acceleration up to relativistic energies (e.g.
\cite{bednarek 2007, eichler-usov 1993}). The clean detection of non-thermal radio
emission coming from the colliding wind region of binary systems like WR140, WR146,
etc., indicates that electrons \cite{bena-romero 2003} and protons \cite{benaglia al
2005} are being efficiently accelerated up to relativistic energies. In Wolf-Rayet / OB
systems, like WR140, luminosities can reach up to 10$^{34}$ erg/s at E > 100 MeV
\cite{pittard-dough 2006}. Finally, supernova explosions of very massive stars in SFR
open clusters can also trigger collective shocks where particles can be accelerated
[e.g. \cite{parizot al 2004}].

The spectral energy distribution can be calculated on the 
same footing as in the XRB systems, taking into account both leptonic and hadronic shock
acceleration and cooling by scattering with matter and radiation. In this case one also
considers bremsstrahlung cooling\footnote{Thermal bremsstrahlung from an ionized
hydrogen cloud (HII region) is often
called free-free emission because it is produced by free electrons scattering off ions
without being captured; the electrons are free before the interaction and remain free
afterwards.}.

\subsubsection{Extragalactic VHE emitters}

Besides galactic SFRs, star-forming galaxies might be also considered energetic
gamma-ray emitters and likely energetic neutrino point sources. In both cases cosmic
rays would be accelerated by the above mentioned mechanisms in order to produce these
high-energy emissions. So far, however, only one extragalactic galaxy (the Large
Magellanic Cloud) was detected with EGRET,  with a dim integral gamma-ray flux
1.9(+/-0.4) x 10$^{-7}$ ph/cm2 s ($E$> 100 MeV) \cite{sreekumar al 1992}.
Thus, based on the $\gamma$-$\nu$ connection, it would demand several years to detect
neutrinos from the LMC using IceCube (unless there was an
anomalous hardening of the spectrum) \cite{dermer 2006}. Nevertheless, the superpositions
of the neutrino emission from star-forming galaxies should grant a background as derived
from the synchrotron radio luminosity associated with cosmic-ray acceleration
\cite{loeb-waxman 2006}. Other likely neutrino sources are gamma-ray bursters and
TeV blazars since their bright and hard gamma-ray spectra could plausibly have a
hadronic origin.

The leading model for gamma-ray bursters involves a relativistic fireball jet, where the gamma rays
are produced from Fermi-accelerated particles in optically thin shocks (for a review
see, e.g., \cite{meszaros 2006}). In the so-called collapsar model for long-duration
gamma-ray bursts (GRBs), the core of a massive star collapses to a black hole or neutron star, driving a
highly relativistic jet which breaks out of the star \cite{woosley 1993}. Within this
framework, high-energy neutrinos from relativistic proton collisions have been studied
(e.g. \cite{waxman-bahcall 1997, alvarez-muniz al 2000}) predicting energies up to $E_{\nu}>
10^5$ TeV from external shocks \cite{waxman-bahcall 2000, dermer-atoyan prl2003}. In
addition, internal shocks can occur while the relativistic jet is still in the star, with
a neutrino precursor burst of $E_{\nu}\geq 5$ TeV emitted from an imprisoned jet
that is dark in gamma rays but bright in neutrinos \cite{meszaros-waxman 2001}.

Present detection rates suggest that low-luminosity GRBs with mildly relativistic
outflows,  like GRB 060218, are two orders of magnitude more common than conventional
GRBs \cite{campana al 2006, waxman al 2007}. In addition, mildly relativistic jets are
supposed to be more baryon rich. Thus, their contribution to the neutrino background
could be even larger than conventional high-luminosity GRBs \cite{gupta-zhang 2007,
ando-beacom 2005}.

Neutrino production in GRBs depends on the Doppler factor of the blast wave and the
baryon load related to the energy contribution of nonthermal protons \footnote{ It has
been argued that with the exception of the signal originating in an initial 'proper'
gamma-ray burst, all the other spikes and time variabilities can be explained by the
interaction of the accelerated-baryonic-matter pulse with inhomogeneities in the
interstellar matter \cite{ruffini al 2001}.}. Provided that the baryon-loading factor is
well above 10 and the Doppler factor is $\leq$ 200, as required to produce ultra
high-energy cosmic rays in the framework of the collapsar model, neutrinos from GRBs
might be detectable with IceCube \cite{dermer-atoyan prl2003}.

Given that particle acceleration up to energies well above the TeV scale must take place
in TeV blazars, it is assumed they are the most probable neutrino sources. However  it
is more likely that the so-called flat spectrum radio blazars are brighter neutrino
sources \cite{atoyan-dermer 2003}. The reason is that some flat spectrum radio quasars
(FSRQs) attain fluxes ten times brighter than the TeV (BL Lac) blazar fluxes. In fact,
the absence of TeV radiation in FSRQs could be just a consequence of opacity, i.e.
$\gamma\gamma$
attenuation with the extragalactic photon background or its own accretion disc
radiation. In a conservative analysis it has been shown that
 IceCube could detect one or several neutrinos during bright FSRQ
blazar flares, such as that observed from 3C 279 in 1996 \cite{atoyan-dermer 2001}.

%



\begin{theacknowledgments}
The author is grateful to G.E. Romero for a critical reading of the manuscript
and useful remarks, and M.M. Reynoso for fruitful discussions about emission processes
in XRBs. Thanks are also due to FUNCAP (Brazil) for financial support.
\end{theacknowledgments}

\bibliographystyle{aipproc}   







\end{document}